# An Economic Analysis of User-Privacy Options in Ad-Supported Services

Joan Feigenbaum[1*], Michael Mitzenmacher[2**], and Georgios Zervas[1***]

[1] Computer Science Department, Yale University
[2] School of Engineering & Applied Sciences, Harvard University

**Abstract.** We analyze the value to e-commerce website operators of offering privacy options to users, *e.g.*, of allowing users to opt out of ad targeting. In particular, we assume that site operators have some control over the cost that a privacy option imposes on users and ask when it is to their advantage to make such costs low. We consider both the case of a single site and the case of multiple sites that compete both for users who value privacy highly and for users who value it less. One of our main results in the case of a single site is that, under normally distributed utilities, if a privacy-sensitive user is worth at least $\sqrt{2}-1$ times as much to advertisers as a privacy-insensitive user, the site operator should strive to make the cost of a privacy option as low as possible. In the case of multiple sites, we show how a Prisoner's-Dilemma situation can arise: In the equilibrium in which both sites are obliged to offer a privacy option at minimal cost, both sites obtain lower revenue than they would if they colluded and neither offered a privacy option.

## 1 Introduction

Advertising supports crucially important online services, most notably search. Indeed, more than 95% of Google's total revenue derives from advertising.[3] Other advertiser-supported websites provide a growing array of useful services, including news and matchmaking. Because of its essential role in e-commerce, online advertising has been and continues to be the subject of intensive study by diverse research communities, including Economics and Computer Science. In this paper, we focus on an aspect of online advertising that has received little attention to date: how website operators can maximize their revenue while permitting privacy-sensitive users to avoid targeted ads.

*Targeted ads* are those chosen to appeal to a certain group of users. In *contextual targeting*, ads are matched to search queries or other commands issued by users; because it does not entail the collection and mining of any information except that which the user provides voluntarily and explicitly at the time the ad is placed, it is not usually viewed as intrusive, and few users try to avoid it. In *demographic targeting*, ads are matched to users' demographic categories such as gender, location, age, race, religion, profession, or income. Demographic targeting may involve considerable data collection and analysis, and it is more controversial than contextual targeting: Some users are uncomfortable about being categorized, feel more vulnerable to ads that make use of their demographic categories, and worry that the same demographic information may be used for purposes more consequential and nefarious than advertising; other users appreciate the fact that their membership in certain demographic categories, particularly age, gender, and location, can prevent their being shown numerous time-wasting ads that are provably irrelevant to them. In *behavioral targeting*, ads are matched to individual users' browsing histories. By definition, it involves the collection and analysis of sensitive information, and many users take steps to avoid it.

---

[*] Supported in part by NSF grant CNS-1016875.
[**] Supported in part by NSF grants IIS-0964473 and CCF-0915922 and by grants from Google Research and Yahoo Research.
[***] Supported by a Simons Foundation Postdoctoral Fellowship.

[3] http://investor.google.com/financial/tables.html

Unsurprisingly, targeted ads are more effective than generic, untargeted ads, and thus they fetch higher prices. Behavioral targeting, for example, has been shown to produce higher click-through rates than no targeting; estimates of the extent of improvement in click-through rates vary widely, however, from 20% in the work of Chen *et al.* [3] to a factor of six in the work of Yan *et al.* [11]. Conversion rate, *i.e.*, the fraction of those users who, after clicking through to the advertiser's site actually buy something, is also higher for targeted ads; see Beales [1] for a discussion of the effect of behavioral targeting on conversion rates and Jansen and Solomon [7] on the effect of demographic targeting in general and gender in particular. Although efforts to quantify the effect of ad targeting on website operators' revenues are ongoing, there is credible evidence that the effect is large enough to imply that the elimination of targeted ads could mean the end of the Web as we know it; Goldfarb and Tucker [4], for example, studied the effect of EU privacy regulation on ad revenue and concluded that, all else equal, advertisers would have to spend approximately $14.8B more annually to achieve the same effectiveness under a strict privacy regime (*i.e.*, one that is less friendly to targeted ads) and that the dependence on targeted ads is highest among general-audience websites, such as those that provide news or weather.

Because targeted ads are lucrative for website operators, users are being observed, categorized, and tracked ever more precisely. Understandably, some users fear loss of privacy, and various tools offered by website operators (*e.g.*, opt-outs and other customizable privacy settings) and by third parties (*e.g.*, anonymizing browser plug-ins such as Torbutton[4]) that promise online-privacy protection are proliferating. These tools allow users to avoid targeted ads, but of course people who use them can still be shown generic, untargeted ads. Although many such tools are available without charge, they can impose non-monetary costs on users, *e.g.*, time and effort spent figuring out the often obscure privacy options presented by a UI, time and effort spent on installation of new software such as a privacy-enhancing browser plug-in, and reduced ease of use, speed, and/or quality of service. To a considerable extent, these costs can be controlled by website operators.

---

[4] See https://www.torproject.org/torbutton/.

We ask when it is to website operators' advantage to make the cost of such privacy options low. Our major contributions include:

– Economic models in which to address the problem, both for the case of a single site and for that of multiple sites that compete both for users who value privacy very highly and for users who value it less.
– A complete analysis of the case of a single site with homogeneous users and normally distributed utilities. In this setting, if a privacy-sensitive user is worth at least 0.4 times as much to advertisers as a privacy-insensitive user, the site operator should strive to make the cost of a privacy option as low as possible.
– A complete analysis of the case of two sites with user demand functions that denote their privacy preferences. In this setting, we show how a Prisoner's-Dilemma situation can arise: In the equilibrium in which both sites are obliged to offer a privacy option at minimal cost, both sites obtain lower revenue than they would if they colluded and neither offered a privacy option.

## 2   Related Work

To the best of our knowledge, we are the first to study the question of when it is to advantage of website operators to minimize the cost of providing users with privacy options. However, several related aspects of web users' ability to control their personal information have been studied.

Riederer *et al.* [8] propose a market for personal information based on the notion of *transactional privacy*. Users decide what information about themselves should be for sale, and aggregators buy access to users' information and use it to decide what ads to serve to each user. The information market that connects users and aggregators, handles payments, and protects privacy achieves truthfulness and efficiency using an unlimited-supply auction.

Carrascal *et al.* [2] use *experience sampling* to study the monetary value that users place on difference types of personal information. They find, for example, that users place a significantly higher value on information about their offline behavior than they do on information about their browsing behavior. Among categories of online information, they value financial and social-network information more highly than search and shopping information.

Iyer, Soberman, and Villas-Boas [6] consider advertising strategies in segmented markets, where competing firms can target ads to different segments. They find that firms can increase profits by targeting more ads at consumers who have a strong preference for their product than at comparison shoppers who might be attracted to the competition. Interestingly, targeted advertising produces higher profits regardless of whether the firms can price discriminate. Moreover, the ability to target advertising can be more valuable to firms in a competitive environment than the ability to price discriminate.

Telang, Rajan, and Mukhopadhyay [10] address the question of why multiple providers of free, online search services can coexist for a long time. In standard models of vertical (or quality) differentiation, a lower-quality product or service must sell for a lower price than its higher-quality competitor if it is to remain in the marketplace; if the prices are equal, all consumers choose the higher-quality alternative. Similarly, in standard models of horizontal (or taste) differentiation, sustained differentiation among products or services occurs when users incur high transportation costs. Yet, neither price nor transportation cost is a strategic variable in online search. Telang, Rajan, and Mukhopadhyay point out that, although the quality of one search service may clearly be higher than that of its competitors *on average*, the quality of results of *a particular search by a particular user* is highly variable and inherently stochastic. Thus, there is a nontrivial probability that a user will be dissatisfied with the results of a particular search and wish to search again for the same information, using a different search service. It is precisely the zero-price, zero-transportation-cost nature of the user's task that may cause him to use more than one search service in a single session. In the aggregate, this feature creates residual demand for lower-quality search services, allowing them to coexist with their higher-quality competitor.

## 3  Single-provider case

We begin by presenting a general model of an ad-supported service with a single provider. Let $n$ be the size of the market for this service (*i.e.*, the number of users), $v$ be the revenue extracted by the service provider for each user that allows targeted ads (referred to below as a "targeted user"), and $\gamma v$, with $0 < \gamma \leq 1$, the revenue extracted for each user that avoids targeted ads (referred to below as a "private user"). The total revenue extracted by the provider is given by:

$$r = nv(s + \gamma p), \qquad (1)$$

where $s$ is the fraction of the market that consists of targeted users, and $p$ is the fraction that consists of private users.

In this setting, we model the users by way of their utilities. For a specific user, the random variables $U^S$ and $U^P$ denote the utilities that the provider derives from the targeted and private services, respectively. We discount the utility by a cost $c$, which can be thought of as fixed cost that the user pays to set up the private option. In our model, we assume $c$ is under the control of the service provider; hence, the provider will choose the cost of the privacy option to optimize revenue.[5] We assume $c \geq 0$, but one could expand our analyses to include settings in which the provider "pays" users to use its service and thereby induces a negative cost $c$. Let $U = (U^S, U^P)$ be the corresponding joint distribution. Let $f(x,y) = \Pr[U^S = x, U^P = y]$ be the joint density, and similarly let $F(x,y) = \Pr[U^S \leqslant x, U^P \leqslant y]$ be the joint distribution function.

---

[5] One could imagine more elaborate models, where the cost was governed by a distribution, and the provider could, for example, control the mean of the distribution; for simplicity, we focus on the constant-cost model in this first paper.

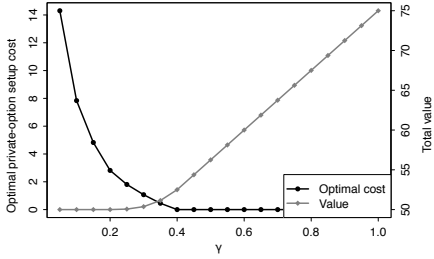 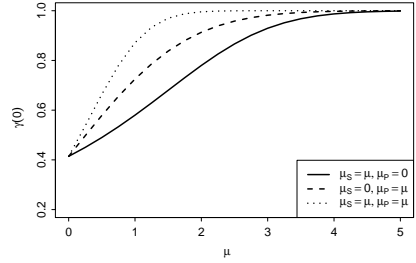

Fig. 1: Optimal privacy-option setup cost and value for various $\gamma$ and $n = 100$, $v = 1$.

Fig. 2: The value of $\gamma(0)$ – where the privacy option takes on zero cost – for three settings: $U^P$'s mean fixed at 0, $U^S$'s mean fixed at 0, and $U^P$ and $U^S$ have the same mean.

A user may:

1. use the targeted option and derive utility $U^S$;
2. use the private option and derive utility $U^P - c$;
3. abstain from using the service for a utility of 0.

Users choose among the above options to maximize their utility. Their choices determine the values of $s$ and $p$.

From the standpoint of the provider, finding the revenue-maximizing cost $c^*$ involves computing trade-offs between $s$ and $p$. We have:

$$s = \Pr[U^P - U^S < c, U^S \geq 0] = \int_0^\infty \int_{-\infty}^{c+y} f(x,y)dxdy, \qquad (2)$$

$$p = \Pr[U^P - U^S \geq c, U^P \geq c] = \int_0^\infty \int_{c+y}^\infty f(x,y)dxdy + \int_{-\infty}^0 \int_c^\infty f(x,y)dxdy. \qquad (3)$$

We emphasize that, in this model, $s + p$ may be less than 1, because users with negative utility from both targeted and private options will not use the service at all.

### 3.1 Normally Distributed User Utilities

We now explore this model by considering the case of normally distributed user utilities. Assume that $U = (U^S, U^P)$ follows a standard bivariate normal distribution with mean vector zero and covariance matrix $\Sigma = \{\{1, \rho\}, \{\rho, 1\}\}$; here, $\rho$ is the correlation coefficient between $U^S$, and $U^P$. Use $\phi_2$ to denote $U$'s density and $\Phi_2$ to denote its distribution function. The marginal distributions $U^S$ and $U^P$ are standard normal with mean 0, variance 1, density function

$\phi(x) = \frac{1}{\sqrt{2\pi}} \exp(-x^2/2)$, and distribution function $\Phi(x) = \frac{1}{2}\left[1 + \text{erf}(x/\sqrt{2})\right]$.
We first consider the case in which $\rho = 0$.

### 3.2 Uncorrelated user utilities, $\rho = 0$.

The fraction of targeted users is

$$s = \Pr[U^P - U^S < c, U^S \geq 0] \tag{4}$$

$$= \int_0^\infty \int_{-\infty}^{c+y} \phi(x)\phi(y) dx dy. \tag{5}$$

Similarly, the fraction of private users is

$$p = \Pr[U^P - U^S \geq c, U^P \geq c] \tag{6}$$

$$= \int_0^\infty \int_{c+y}^\infty \phi(x)\phi(y) dx dy + \int_{-\infty}^0 \int_c^\infty \phi(x)\phi(y) dx dy, \tag{7}$$

where $\text{erfc}(x) = 1 - \text{erf}(x)$. Observe that $s$ is monotonically increasing in $c$, while $p$ is monotonically decreasing in $c$. The rate of change of the fraction of targeted users as a function of $c$ is:

$$\frac{\partial s}{\partial c} = \int_0^\infty \phi(y)\phi(c+y)\,dy \tag{8}$$

$$= \int_0^\infty \frac{e^{-\frac{y^2}{2}}}{\sqrt{2\pi}} \frac{e^{-\frac{1}{2}(c+y)^2}}{\sqrt{2\pi}}\,dy \tag{9}$$

$$= \int_0^\infty \frac{e^{-\frac{1}{2}(c+y)^2 - \frac{y^2}{2}}}{2\pi}\,dy \tag{10}$$

$$= \frac{e^{-\frac{c^2}{4}} \text{erfc}\left(\frac{c}{2}\right)}{4\sqrt{\pi}}, \tag{11}$$

which is easily seen to be decreasing in $c$. Using similar calculations, we can compute the rate of change of the fraction of private users with respect to $c$:

$$\frac{\partial p}{\partial c} = \int_{-\infty}^0 -\phi(c)\phi(y)\,dy + \int_0^\infty -\phi(y)\phi(c+y)\,dy \tag{12}$$

$$= -\frac{e^{-\frac{c^2}{2}}\left(e^{\frac{c^2}{4}} \text{erfc}\left(\frac{c}{2}\right) + \sqrt{2}\right)}{4\sqrt{\pi}}, \tag{13}$$

which is similarly increasing in $c$. The provider is indifferent with respect to revenue earned between the two types of users when:

$$\frac{\partial s}{\partial c} = -\gamma \frac{\partial p}{\partial c}. \tag{14}$$

Denote by $\gamma(c)$ the value of $c$ for which equality holds. Substituting and solving for $\gamma(c)$, we obtain:
$$\gamma(c) = 1 - \frac{2}{\sqrt{2}e^{\frac{c^2}{4}}\operatorname{erfc}\left(\frac{c}{2}\right) + 2}. \tag{15}$$

We continue by proving an auxiliary lemma.

**Lemma 1.** *$\gamma(c)$ in decreasing in $c$.*

*Proof.* Consider the derivative
$$\frac{\partial e^{\frac{c^2}{4}}\operatorname{erfc}\left(\frac{c}{2}\right)}{\partial c} = \frac{1}{2}ce^{\frac{c^2}{4}}\operatorname{erfc}\left(\frac{c}{2}\right) - \frac{1}{\sqrt{\pi}}. \tag{16}$$

We show that it is negative for all $c > 0$; this suffices to prove the lemma. Note the following equivalences.
$$\frac{1}{2}ce^{\frac{c^2}{4}}\operatorname{erfc}\left(\frac{c}{2}\right) - \frac{1}{\sqrt{\pi}} < 0 \tag{17}$$
$$ce^{\frac{c^2}{4}}\operatorname{erfc}\left(\frac{c}{2}\right) < \frac{2}{\sqrt{\pi}} \tag{18}$$
$$ce^{\frac{c^2}{4}}\frac{2}{\sqrt{\pi}}\int_{c/2}^{\infty} e^{-t^2}\,dt < \frac{2}{\sqrt{\pi}} \tag{19}$$
$$ce^{\frac{c^2}{4}}\int_{c/2}^{\infty} e^{-t^2}\,dt < 1. \tag{20}$$

We prove the last line above, using the following bound of [9] (a tighter version of Komatsu's inequality [5]):
$$e^{x^2}\int_{x}^{\infty} e^{-t^2}\,dt < 2/\left(3x + \sqrt{x^2 + 4}\right). \tag{21}$$

At $x = c/2$, we obtain
$$e^{\frac{c^2}{4}}\int_{c/2}^{\infty} e^{-t^2}\,dt < 2/\left(3c/2 + \sqrt{c^2/4 + 4}\right), \tag{22}$$

which yields
$$ce^{\frac{c^2}{4}}\int_{c/2}^{\infty} e^{-t^2}\,dt < 2c/\left(3c/2 + \sqrt{c^2/4 + 4}\right) < 1, \tag{23}$$

proving the lemma. □

We have that $\gamma(0) = \sqrt{2} - 1 \approx 0.41$. Furthermore, because $\gamma(c)$ is decreasing in $c$, for any $\gamma \geq \gamma(0)$, it follows that the provider's best strategy is $c = 0$, *i.e.*, offering a free privacy option. Using the above, we are now ready to state the main theorem of this section. corollary

**Theorem 1.** *If $U$ follows a standard bivariate normal distribution with correlation $\rho = 0$, then the provider will offer a free privacy option whenever $\gamma \geq \sqrt{2}-1$.*

*Remark 1.* The specific value $\sqrt{2}-1$ arises from our assumption that $U^P$ and $U^S$ were distributed according to a standard normal distribution of mean 0. Similar results for other means and variances can be calculated in a similar fashion. Figure 2 demonstrates three variations: where the mean of $U^P$ is fixed at 0 but the mean of $U^S$ varies; where the mean of $U^S$ is fixed at 0 but the mean of $U^P$ varies; and where the means vary but are equal. (All variances remain 1.) For example, where $U^S$ and $U^P$ have equal means, we see that $\gamma(0)$ converges to 1 very quickly, as offering privacy cannibalizes more lucrative targeted users more readily than it garners new private users.

### 3.3 Correlated utilities, $\rho \neq 0$.

Assume that $U = (U^S, U^P)$ follows a standard bivariate normal distribution with correlation $\rho$. Use $\phi_2$ to denote its density function. We derive an indifference condition similar to the one in Equation 14:

$$\gamma \left( \int_{-\infty}^{0} \phi_2(c, y) \, dy + \int_{0}^{\infty} \phi_2(c+y, y) \, dy \right) = \int_{0}^{\infty} \phi_2(c+y, y) \, dy. \quad (24)$$

Substituting for the density function of the standard bivariate normal and integrating, we obtain an expression for $\gamma$ in terms of $c$ and $\rho$:

$$\gamma = \left( \frac{\sqrt{2-2\rho} \, e^{\frac{c^2(1-2\rho)}{4(\rho-1)}} \operatorname{erfc}\left(\frac{c\rho}{\sqrt{2-2\rho^2}}\right)}{\operatorname{erfc}\left(\frac{c}{2\sqrt{\rho+1}}\right)} + 1 \right)^{-1}. \quad (25)$$

As before, setting $c = 0$ yields

$$\gamma(0) = \frac{1}{1+\sqrt{2-2\rho}}. \quad (26)$$

We observe that, as the correlation coefficient increases (reps., decreases) the value of $\gamma$ beyond which it makes sense to offer a privacy option at no cost also increases (reps., decreases). That is, greater correlation means one requires higher revenue from private users in order to offer privacy at no cost; in particular, when $U^P = U^S$, Equation 26 reasonably requires that $\gamma(0) = 1$. We can now state a generalized version Theorem 1, taking into account correlated user utilities.

**Theorem 2.** *If $U$ follows a standard bivariate normal distribution with correlation $\rho$, then the provider will offer a free privacy option whenever $\gamma \geq \frac{1}{1+\sqrt{2-2\rho}}$.*

## 4 A two-player game

We provide a general model of the two-player version of the game. We then explore the model by delving into a concrete example. Throughout this section, we use the terms "player" and "provider" interchangeably. As in the single-provider case, a "targeted user" is one who does not use the privacy option, and a "private user" is one who does.

The game proceeds in two periods $t = 1, 2$. To begin, at $t = 1$, we have two providers $S_i$, for $i = 1, 2$, that offer competing, advertising-supported, non-private services. We let $S_0$ denote a user's outside option, which in this case is to use neither service. Denote the fraction of users who choose $S_i$ at time $t$ by $s_{it}$. We have $\sum_{i=0}^{2} s_{i1} = 1$.

At $t = 2$, simultaneously, both providers can introduce private variants, *i.e.*, ones in which users avoid targeted ads; we denote these by $P_i$. The providers can determine an associated "cost" that controls the utility of the private variants, with the goal of tuning the market share for each service they provide. We denote these costs by $c_i$. The fraction of users that choose $P_i$ at time $t$ is given by $p_{it}$. We have $\sum_{i=0}^{2} s_{i2} + \sum_{j=1}^{2} p_{j2} = 1$. The fraction of users left using the non-private options (or neither option) at $t = 2$ is given by:

$$s_{i2} = s_{i1}(1 - F_i(c_1, c_2)) \text{ for } i = 0, 1, 2.$$

That is, $F_i(c_1, c_2)$ is the fraction of users who were using $S_i$ or were not using either service but are now using one of the private variants. The users $s_{i1} - s_{i2}$ switching to a private variant are distributed among the two providers as follows:

$$p_{i2} = H_i(c_1, c_2) \sum_{j=0}^{2} (s_{j1} - s_{j2}) \text{ for } i = 1, 2.$$

Here, $H_i(c_1, c_2)$ is a function determining the split, among the competing private-service providers, of users switching to a private variant; note $H_1(c_1, c_2) = 1 - H_2(c_1, c_2)$. Also, recall that $p_{i1} = 0$.

Let $v_i$ be the value that provider $i$ derives from the standard service. Let $\gamma_i v_i$ be the value it derives from the private service, where $\gamma_i \in (0, 1]$. Let $r_i$ denote the revenue function of provider $i$ at the second stage of the game. We have:

$$\begin{aligned}
r_i &= v_i s_{i2} + v_i \gamma_i p_{i2} \\
&= v_i s_{i1}(1 - F_i(c_1, c_2)) + v_i \gamma_i H_i(c_1, c_2) \sum_{j=0}^{2}(s_{j1} - s_{j2}) \\
&= v_i s_{i1}(1 - F_i(c_1, c_2)) + v_i \gamma_i H_i(c_1, c_2) \sum_{j=0}^{2}(s_{j1} - s_{j1}(1 - F_j(c_1, c_2))) \\
&= v_i \left( s_{i1}(1 - F_i(c_1, c_2)) + \gamma_i H_i(c_1, c_2) \sum_{j=0}^{2} s_{j1} F_j(c_1, c_2) \right)
\end{aligned}$$

(27)

for $i = 1, 2$.

In equilibrium, neither provider can increase its revenue by unilaterally deviating. The first-order conditions (FOC) are given by

$$\frac{\partial r_i}{\partial c_i} = 0, \text{ for } i = 1, 2. \tag{28}$$

Let $\{\hat{c}_1, \hat{c}_2\}$ be a solution to this system of equations. Then, the second-order conditions (SOC) are given by:

$$\frac{\partial^2 r_i}{\partial c_i^2}(\hat{c}_1, \hat{c}_2) < 0, \text{ for } i = 1, 2. \tag{29}$$

Our questions revolve around the equilibrium of this game.

### 4.1 A prisoners' dilemma

The framework described above was designed to be very general; however, this makes it somewhat difficult to get a handle on the nature of the game. We consider a worked example in order to gain more insight. For simplicity, we will assume $F_i = F$ for $i = 0, 1, 2$. There are certain natural properties that we want for the function $F$: $F$ should be decreasing in the $c_i$, and $F$ should go to 0 as both $c_i$ go to infinity. We simply things further by assuming that, if either cost $c_i$ goes to 0, all users will prefer the zero-cost private option, and thus $F$ will go to 1. This may not be the case in all settings, but it is a reasonable and instructive place to start.

A relatively straightforward function with these properties is

$$F(c_1, c_2) = \begin{cases} 0 & \text{if } c_1 = c_2 = \infty, \\ 1 & \text{if } c_1 = 0, \text{ or } c_2 = 0, \\ \exp(-\frac{c_1 c_2}{c_1 + c_2}) & \text{otherwise.} \end{cases} \tag{30}$$

We define $H_i$ so that the fraction that goes to $P_i$ is proportional to its cost.

$$H_1(c_1, c_2) = \begin{cases} 0 & \text{if } c_1 = \infty, \\ \frac{1}{2} & \text{if } c_1 = c_2 = 0, \\ \frac{c_2}{c_1 + c_2} & \text{otherwise.} \end{cases} \tag{31}$$

We define $H_2$ similarly. Also, for notational simplicity let $s_{i1} = s_i$. Under these assumptions, the payoff function for player 1 is:

$$r_1 = v_1 \left( \frac{e^{-\frac{c_1 c_2}{c_1 + c_2}} (c_2 \gamma_1 - (c_1 + c_2) s_1)}{c_1 + c_2} + s_1 \right), \tag{32}$$

and similarly, for player 2. Note that the payoffs are continuous. Furthermore, under the assumption that $F_i = F$, the first and second derivatives of the revenue function assume the following forms:

$$\frac{\partial r_i}{\partial c_i} = v_i \left( -s_{i1} \frac{\partial F}{\partial c_i} + \gamma_i \frac{\partial H_i}{\partial c_i} F(c_1, c_2) + \gamma_i H_i(c_1, c_2) \frac{\partial F}{\partial c_i} \right), \tag{33}$$

and
$$\frac{\partial^2 r_i}{\partial c_i^2} = v_i \left( -s_{i1} \frac{\partial^2 F}{\partial c_i^2} + \gamma_i \left( 2 \frac{\partial H_i}{\partial c_i} \frac{\partial F}{\partial c_i} + \frac{\partial^2 H_i}{\partial c_i^2} F + \frac{\partial^2 F}{\partial c_i^2} H \right) \right). \quad (34)$$

## 4.2 Computation of equilibria

We now assume that $\gamma_i > 0$, i.e., that both players can derive some revenue from private users.

**Theorem 3.** *The game has two possible equilibria:*

1. *at* $\{c_1^* = 0, c_2^* = 0\}$, *with* $r_i = v_i \gamma_i / 2$,
2. *at* $\{c_1^* = \infty, c_2^* = \infty\}$, *if* $s_i > \gamma_i$, *with* $r_i = s_i v_i$.

*Remark 2.* Theorem 3 demonstrates how the Prisoners' Dilemma arises naturally in this two-player game. For example, if $s_i = \frac{3}{4}\gamma_i$ for $i = 1, 2$, then, when neither provider offers a privacy option, their revenue is $\frac{3}{4}\gamma_i v_i$. However, this is not an equilibrium point; at equilibrium, both players offer zero-cost privacy options, and their revenue is reduced to $\frac{1}{2}\gamma_i v_i$.

*Proof.* The proof proceeds in a sequence of lemmas. We begin by considering cases in which one player offers a free privacy option, while the other charges a (possibly infinite) cost.

**Lemma 2.** *The game does not admit solutions of the form* $\{c_i = 0, c_{-i} > 0\}$.

*Proof.* Because the game is symmetric, it suffices to consider the case in which $c_1 = 0$, and $c_2 = c$, for some constant $c > 0$. From Equations 30 and 31, we have $F(0, c) = 1$, $H_1(0, c) = 1$, and $H_2(0, c) = 0$. From Equation 27, the revenue for player 1 is $r_1 = v_1 \gamma_1$, and for player 2 it is $r_2 = 0$. Suppose player 2 unilaterally deviates by playing $c_2 = 0$. In this case, $H_2(0,0) = 1/2$, and $r_2 = v_2 \gamma_2 / 2$. Therefore, because $\gamma_2 > 0$, $c_1 = 0, c_2 = c$, does not constitute an equilibrium. □

Next, we consider settings in which both players offer the privacy option for a finite, non-zero cost.

**Lemma 3.** *The game does not admit solutions of the form* $\{c_i > 0, c_{-i} > 0\}$.

*Proof.* We consider candidate equilibria suggested by solutions to the FOC:

$$F(c_1, c_2) \frac{c_2 v_1 \left( c_2 (c_1 + c_2) s_1 - (c_2^2 + c_2 + c_1) \gamma_1 \right)}{(c_1 + c_2)^3} = 0 \quad (35)$$

$$F(c_1, c_2) \frac{c_1 v_2 \left( c_1 (c_1 + c_2) s_2 - (c_1^2 + c_1 + c_2) \gamma_2 \right)}{(c_1 + c_2)^3} = 0. \quad (36)$$

First, note that, because costs are finite, $F(c_1, c_2) > 0$. Therefore, the FOC can be simplified to the following equivalent conditions:

$$c_2 (c_1 + c_2) s_1 - (c_2^2 + c_2 + c_1) \gamma_1 = 0, \quad (37)$$
$$c_1 (c_1 + c_2) s_2 - (c_1^2 + c_1 + c_2) \gamma_2 = 0. \quad (38)$$

Solving the first equation above for $c_1$ and substituting into the second one, we obtain the following solution:

$$\left\{c_1 = \frac{\gamma_2}{s_2 - \gamma_2}K, c_2 = \frac{\gamma_1}{s_1 - \gamma_1}K\right\}, \tag{39}$$

where

$$K = \frac{\gamma_1 s_2 + \gamma_2 s_1 - 2\gamma_1\gamma_2}{\gamma_1 s_2 + \gamma_2 s_1 - \gamma_1\gamma_2}. \tag{40}$$

Next, we need to check the SOC for this solution. A long sequence of calculations yields:

$$\frac{\partial^2 r_1}{\partial c_1^2}(c_1, c_2) = \frac{\gamma_1^4 v_1 e^{-\frac{\gamma_1\gamma_2}{\gamma_1 s_2 + \gamma_2(s_1 - \gamma_1)}}(s_1 - \gamma_1)(s_2 - \gamma_2)^4(\gamma_1 s_2 + \gamma_2 s_1 - \gamma_1\gamma_2)}{(\gamma_1 s_2 + \gamma_2 s_1 - 2\gamma_1\gamma_2)^5} < 0,$$

$$\frac{\partial^2 r_2}{\partial c_2^2}(c_1, c_2) = \frac{\gamma_2^4 v_2 e^{-\frac{\gamma_1\gamma_2}{\gamma_1 s_2 + \gamma_2(s_1 - \gamma_1)}}(s_2 - \gamma_2)(s_1 - \gamma_1)^4(\gamma_1 s_2 + \gamma_2 s_1 - \gamma_1\gamma_2)}{(\gamma_1 s_2 + \gamma_2 s_1 - 2\gamma_1\gamma_2)^5} < 0,$$

which can be further simplified to the equivalent conditions

$$(s_1 - \gamma_1)\frac{\gamma_1 s_2 + \gamma_2 s_1 - \gamma_1\gamma_2}{\gamma_1 s_2 + \gamma_2 s_1 - 2\gamma_1\gamma_2} = \frac{s_1 - \gamma_1}{K} < 0, \tag{41}$$

$$(s_2 - \gamma_2)\frac{\gamma_1 s_2 + \gamma_2 s_1 - \gamma_1\gamma_2}{\gamma_1 s_2 + \gamma_2 s_1 - 2\gamma_1\gamma_2} = \frac{s_2 - \gamma_2}{K} < 0. \tag{42}$$

Notice that $(s_1 - \gamma_1)/K$ has the same sign as $c_1$ in Equation 39. Similarly, $(s_2 - \gamma_2)/K$ has the same sign as $c_2$. Because costs are non-negative, the second-order conditions are not met; this solution minimizes rather than maximizes the revenue of the players and is therefore not an equilibrium point. □

Next, we consider the case in which both players offer free privacy options.

**Lemma 4.** *Both players' offering free privacy options (i.e., $c_1 = c_2 = 0$) constitutes an equilibrium of the game.*

*Proof.* In this case, users switch *en masse* to the private services and are distributed equally between the two providers. The revenue for player $i$ is $v_i\gamma_i/2$. Furthermore, if a player unilaterally deviates and switches to a non-zero cost for privacy, his revenue instantly collapses to zero. Therefore, $c_1 = c_2 = 0$ constitutes an equilibrium to the game. □

Finally, we consider the case in which neither player offers a privacy option.

**Lemma 5.** *Neither player's offering a privacy option (i.e., $c_1 = c_2 = \infty$) constitutes an equilibrium of the game if $s_i < \gamma_i$, for $i = \{1, 2\}$.*

*Proof.* Suppose that neither player offers a privacy option; so $r_i = v_i s_i$. Now, consider the case in which player 1 wishes to deviate unilaterally. (The case for player 2 can be argued identically.) First, note that $\lim_{c_2 \to \infty} F(c_1, c_2) =$

$\exp(-c_1)$, and $\lim_{c_2 \to \infty} H_1(c_1, c_2) = 1$. Therefore, if player 2 doesn't offer a privacy option, *i.e.*, $c_2 = \infty$, player 1 deviates and plays $c_1$; his new revenue will be

$$r_1 = s_1 v_1 (1 - \exp(-c_1)) + v_1 \gamma_1 \exp(-c_1) = s_1 v_1 + v_1 \exp(-c_1)(\gamma_1 - s_1). \quad (43)$$

If $\gamma_1 < s_1$, then player 1 cannot improve his position; therefore, not offering a privacy option constitutes an equilibrium. If $\gamma_1 > s_1$, player 1 can increase his revenue by decreasing his cost; therefore, not offering a privacy option is not an equilibrium. (If $\gamma_1 = s_1$, player 1 cannot strictly improve his revenue by deviating, and we have a weak equilibrium.) □

This concludes the proof of Theorem 3. □

This example demonstrates what we see as the potentially natural outcomes of this two-player dynamic that would extend to three or more players as well. It is possible that no service provider is incentivized to offer a privacy option in this game. In such a setting, one might expect a new entrant to enter the "game" and disrupt the status quo by offering a suitable privacy option, potentially forcing other providers to do so as well. It is also possible that all competitors are inclined to offer a privacy option. In our example, this led to all users' opting to offer privacy, but we could have utilized arguably more realistic functions $F$ to account for the fact that some users might find more value in not offering privacy or might find the cost of doing so non-trivial regardless of the efforts they exert to drive the cost down. Under such settings, we would expect other equilibria to arise; in our example, there were other points at which the first-order conditions but not the second-order conditions were met, but, more generally (for other functional relationships), we could have other equilibria.

## 5 Conclusion

The results of Sections 3 and 4 suggest that website operators could have their cake and eat it, too. By carefully controlling the cost to users of opting out of targeted ads, they could maximize their revenue and respect their users' privacy concerns. Putting this approach into practice would require surmounting at least two major obstacles.

First, a service provider would need a good estimate of the parameter $\gamma$, *i.e.*, the fraction of the revenue derived from a targeted user that can be derived from a private user. The value of $\gamma$ is closely related to the extent to which click-through and conversion rates are improved by various forms of ad targeting, which in turn is still the subject of intensive, ongoing research.

Second, the service provider would need to translate the abstract "cost" $c_i$ of our economic analysis into a concrete privacy-enforcement tool that can be installed and used at a cost of $c_i$. It may suffice to be able to choose between two technological options, one of which is clearly more costly to users than the other, but even this would be nontrivial given the state of the art of privacy enforcement.